\documentclass[%
reprint,
superscriptaddress,
amsmath,amssymb,
aps,
pra, showkeys
]{revtex4-2}

\usepackage{xr-hyper}

\makeatletter
\newcommand*{\addFileDependency}[1]{
	\typeout{(#1)}
	\@addtofilelist{#1}
	\IfFileExists{#1}{}{\typeout{No file #1.}}
}
\makeatother

\newcommand*{\myexternaldocument}[1]{%
	\externaldocument{#1}%
	\addFileDependency{#1.tex}%
	\addFileDependency{#1.aux}%
}
\myexternaldocument{Supplementary}

\usepackage{color}
\usepackage{physics}
\usepackage{graphicx}
\graphicspath{{Figures/}}
\usepackage{dcolumn}
\usepackage{bm}
\usepackage{hyperref}
\hypersetup{bookmarksnumbered, pdfpagemode=UseOutlines, 
colorlinks=true, citecolor=blue, filecolor=blue, linkcolor=blue, urlcolor=blue}

\begin{document}

\preprint{APS/123-QED}

\title{Tunable multi-bit stochasticity in \texorpdfstring{La\textsubscript{0.67}Sr\textsubscript{0.33}MnO\textsubscript{3}}{La0.67Sr0.33MnO3}-based probabilistic bits} 

\author{Ishitro Bhaduri}
\thanks{Author contributed equally}
 \email{Contact author: i.bhaduri@rug.nl}
\affiliation{Zernike Institute for Advanced Materials, University of Groningen, 9747 AG, Groningen, The Netherlands}
\affiliation{Groningen Cognitive Systems and Materials Center, University of Groningen, 9747 AG, Groningen, The Netherlands}
\author{Azminul Jaman}
\thanks{Author contributed equally}
 \email{Contact author: azminul.jaman@rug.nl}
\affiliation{Zernike Institute for Advanced Materials, University of Groningen, 9747 AG, Groningen, The Netherlands}
\affiliation{Groningen Cognitive Systems and Materials Center, University of Groningen, 9747 AG, Groningen, The Netherlands}
\author{Walter Quiñonez}
\affiliation{Instituto de Nanociencia y Nanotecnología, CONICET-CNEA-CAC, 1650, Buenos Aires, Argentina}
\author{Ayush Gupta}
\affiliation{Zernike Institute for Advanced Materials, University of Groningen, 9747 AG, Groningen, The Netherlands}
\author{Tamalika Banerjee}
 \email{Contact author: t.banerjee@rug.nl}
\affiliation{Zernike Institute for Advanced Materials, University of Groningen, 9747 AG, Groningen, The Netherlands}
\affiliation{Groningen Cognitive Systems and Materials Center, University of Groningen, 9747 AG, Groningen, The Netherlands}


\begin{abstract}
One promising approach to combat the rapidly escalating computational demands is to use networks of naturally stochastic units called probabilistic bits (p-bits). To date, hardware implementations of p-bits have predominantly relied on thermally unstable nanomagnets. However, the search continues for alternative material platforms exhibiting easily accessible, intrinsic forms of stochasticity. In this work, we demonstrate hitherto unreported p-bit functionality in epitaxial thin films of La\textsubscript{0.67}Sr\textsubscript{0.33}MnO\textsubscript{3}, a material showing an electrically triggered metal-insulator transition, grown on twin-textured LaAlO\textsubscript{3} substrates. By leveraging the combined effects of phase and structural inhomogeneities, we show two distinct modes of voltage-tunable stochastic operation: clocked binary switching and unclocked multi-bit switching. Our findings suggest that the differences result from variations in the energy landscape near the phase transition. This tunability of the energy landscape is promising for designing diverse stochastic behavior within the same material, highlighting its potential for applications in true random number generation for cryptography and probabilistic computing.

\end{abstract}


\maketitle


\section{Introduction}
In recent years, the computing demands and energy consumption of modern artificial intelligence algorithms have surged dramatically, driving the development of alternative, domain-specific computing paradigms beyond the conventional von-Neumann architecture \cite{Big_data, Beyond_von_Neumann}. One such example is probabilistic computing, often regarded as the classical counterpart of quantum computing \cite{Kaiser, Camsari, Chowdhury}. This computing scheme can efficiently address tasks typically performed by quantum computers --- such as inference \cite{Ostwal2}, invertible logic \cite{Camsari3} and combinatorial optimization \cite{Sutton, Borders} --- while circumventing the associated practical challenges of decoherence and cryogenic operation \cite{Preskill2018}. The fundamental building block of probabilistic computers is the probabilistic bit (p-bit), a stochastic unit which function as a tunable random number generator. The p-bit's normalized output randomly fluctuates between 0 and 1 and can be controlled through an input parameter. Thus, a physical p-bit requires a material platform that is inherently stochastic yet controllable.

Designing hardware implementations of p-bits relies on tuning the energy landscape in materials. While a range of non-CMOS-based physical systems have demonstrated p-bit functionality \cite{Studholme2024, Valle2022, Woo2022, Gutierrez-Finol2023, Park2022, Rhee2023, Liu, Woo2024}, the most widely adopted approach involves thermally unstable nanomagnets, integrated into magnetic tunnel junctions (MTJs) \cite{Camsari2, Ostwal, Borders, Vodenicarevic, Daniels2020, Selcuk2024}. In these systems, a precisely tuned low-energy barrier enables the magnetization to stochastically switch between two states, assisted by thermal fluctuations \cite{Camsari, Debashis}. MTJs with thermally stable magnets have also been implemented as p-bits, where the stochasticity can be induced by voltage-controlled magnetic anisotropy \cite{Shao2023}, spin-orbit \cite{Li2023, Kim2015, Puliafito2020} and spin-transfer torque \cite{Fukushima, Choi} and dual-biasing \cite{Lv2019, Lv2022} (a detailed review can be found in \cite{Zink2022}). Although such MTJ-based solutions are fast, robust and scalable, they are constrained by the complexity of their stack structures and the limited tunability of the constituent materials. These limitations underscore the importance of exploring alternative materials exhibiting natural unpredictability that can be harnessed and controlled in a simple manner.

Strongly correlated oxides like La\textsubscript{0.67}Sr\textsubscript{0.33}MnO\textsubscript{3} (LSMO), with several closely coupled degrees of freedom, are a promising class of materials for this purpose. LSMO exhibits a temperature-dependent metal-insulator transition (MIT) which can be electrically triggered, resulting in volatile resistive switching \cite{Salev2021, Jaman2023, Jaman2025}. In the critical region near the phase transition, the coexistence of multiple competing phases gives rise to rich electrical nonlinearities, which can inherently produce a stochastic response \cite{Imada1998, Dagotto2001}.

In this work, we demonstrate stochastic electrical switching in epitaxially grown thin films of LSMO on twin-textured substrates of LaAlO\textsubscript{3} (LAO). We have exploited the natural phase inhomogeneity in LSMO, coupled with the unique topography of the substrate, to design a cascade of competing and coexisting energy states that the system can probabilistically switch between. The innate structural variability in the substrate allows us to demonstrate two distinct modes of voltage-controlled stochastic operation: clocked binary switching between two current states in one sample, and unclocked multi-bit switching between multiple current states in another. This substrate-induced tunability of the energy landscape provides an avenue for engineering diverse forms of stochasticity within the same material platform. Autocorrelation analyses further confirm that the switching sequences generated by our samples are indeed purely stochastic, underscoring their potential to be used as true random number generators for cryptography and probabilistic computing.

\section{Thin film growth and structural characterization}
In this work, we investigate LSMO films grown on two different LAO substrates. LAO substrates are naturally endowed with twin domains and the equilibrium domain configuration varies from one substrate to another. This is brought to light in Fig. S1, which shows optical microscopy images of the two samples (named L1 and L2), with one of them having a more sparse distribution of twin domains than the other. The LAO substrates of 5$\cross$5$\cross$0.5 mm\textsuperscript{3} are obtained from CrysTec, GmbH, Germany. An atomic force microscopy (AFM) scan of a pristine substrate is shown in Fig. S2(c), displaying the structural twin domains characteristic of LAO. On each twin domain, the surface roughness has a root-mean-square value of 135 pm, indicating a smooth, atomically flat surface. 

Thin films of LSMO, 10 nm thick, are grown on LAO substrates by pulsed laser deposition (PLD) at a substrate temperature of 750 $^{\circ}$C and an oxygen pressure of 0.35 mbar, using a KrF laser with a laser fluence of approximately 2 J\hspace{0.05cm}cm\textsuperscript{-2}. Following growth, the films are annealed at an oxygen pressure of approximately 100 mbar for an hour, after which, they are cooled down to room temperature at a rate of 15 $^{\circ}$C/min. The surface structure of the films is monitored \textit{in situ} during growth using reflective high-energy electron diffraction (RHEED). The RHEED intensity oscillations, shown in Fig. S2(b), indicate layer-by-layer growth. The grown films are epitaxial, atomically flat and highly crystalline, as confirmed by post-growth AFM and x-ray diffraction (XRD) scans, shown in Figs. S2(a) and (c).

\section{Results}
\begin{figure*}[t]
\centering
     \includegraphics[width=0.95\textwidth]{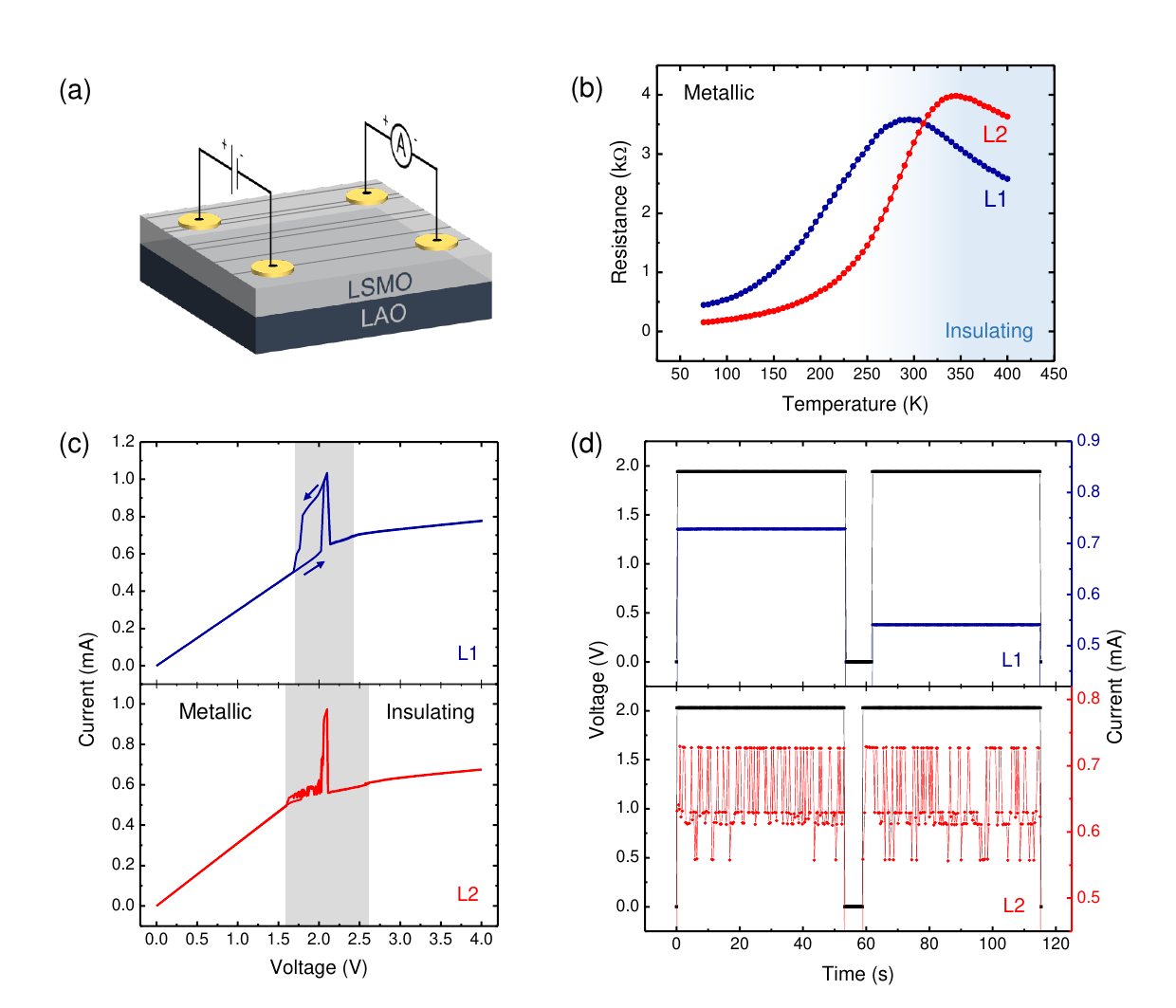}
     \caption{(a) Schematic of the LSMO/LAO samples with the van der Pauw contacts and the electrical connections. The structural impression of the twins is depicted on the LSMO surface. (b) Resistance-temperature curve, showing a temperature-dependent metal-insulator transition (MIT) for both samples. (c) Current-voltage curves, exhibiting a clear slope change around 2 V, demonstrating that the MIT can be electrically triggered. The critical region between two stable states is highlighted in gray. (d) Time-evolution of the current when two constant voltage pulse are applied. The pulse width is approximately 55 s with a delay time of 10 s between the two pulses for both samples. The pulse amplitude is 1.94 V for L1 and 2.03 V for L2. The current axis has been scaled to highlight the switching with more clarity. All measurements have been performed at 295 K.}
     \label{Fig_1}
\end{figure*}

Electrical characterization is done in a van der Pauw geometry using four Ti/Au electrodes, as depicted in Fig.\ref{Fig_1}(a). In this configuration, two electrodes source a DC voltage (or current) and the other two sense the resultant current (or voltage drop). Fig.\ref{Fig_1}(b) shows the temperature-dependence of the electrical resistance. Both LSMO films exhibit a characteristic temperature-dependent MIT. This electrical transition is strongly correlated, via the double-exchange interaction, to the magnetic transition that the material undergoes: from a ferromagnetic state at low temperatures to a paramagnetic state at high temperatures \cite{Tokura1999, Dorr2006, Anderson1955}. The phase transition temperature, \textit{T\textsubscript{MIT}}, determined in Fig. S3, is 235 K for L1 and 290 K for L2.

It has recently been shown that the MIT in LSMO can be electrically triggered via Joule heating \cite{Jaman2023, Salev2021, Jaman2025, Salev2023a, Salev2023b, Salev2024, Chen2024}. In this work, we realize this by performing voltage sweeps while simultaneously measuring the current. Fig. \ref{Fig_1}(c) shows the current, \textit{I}, as a function of voltage, \textit{V}, for the two samples. Both \textit{I-V} curves exhibit a marked slope change near 2 V, concomitant with a transition from a metallic to an insulating state. This resistive switching is volatile, with the system reverting to its initial state at 0 V. Interestingly, close to the transition point, several nonlinear features appear in the \textit{I-V}s. The emergence of such nonlinearities can be associated to a metastable critical state near the transition point, where multiple phases coexist \cite{Dagotto2001, Imada1998}. 
A noteworthy observation is that the exact nature of the critical region differs for L1 and L2. Prior to the slope change, both samples exhibit an abrupt current spike immediately followed by a sharp decrease. For L1, however, the retrace current follows a different path, resulting in a small hysteresis loop while for L2, there appears to be a region where the current randomly switches between multiple values.

\begin{figure*}[t]
\centering
     \includegraphics[width=0.95\textwidth]{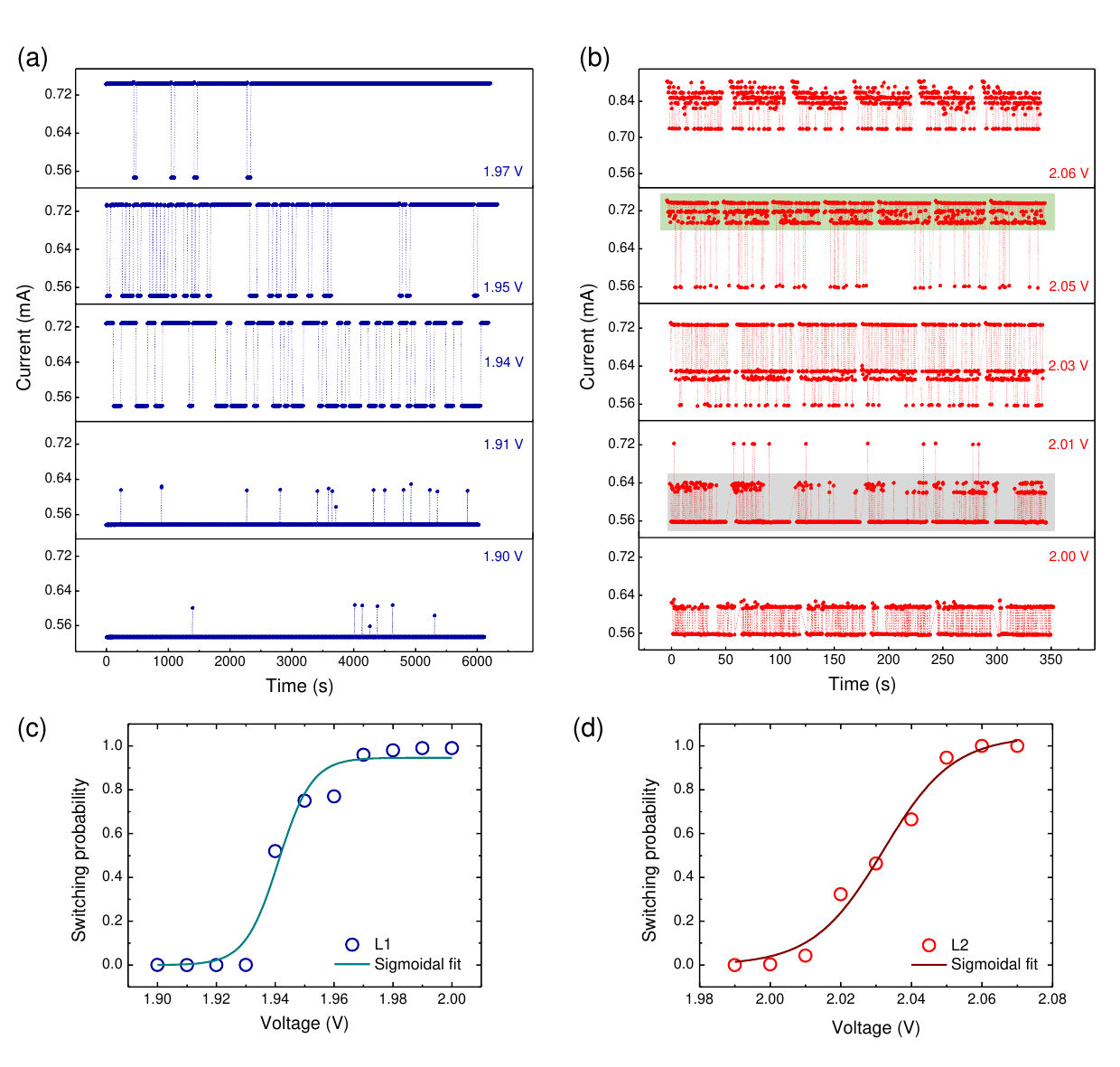}
     \caption{Time-evolution of the current when constant voltage pulses are applied to (a) L1 and (b) L2. 100 pulses are shown for L1 and 6 pulses are shown for L2. Each voltage pulse is applied for approximately 55 s with a 10 s delay between consecutive pulses. L1 mostly switches between two distinct current states while L2 switches between two current regimes --- the low-current regime is highlighted in gray and the high-current regime in green. The probability of switching to the higher current state (or regime) as a function of the voltage pulse amplitude is plotted for (c) L1 and (d) L2, along with a sigmoidal fit to the data. In order to obtain a two-state probability distribution, the time-dependent current values for both samples are binarized with respect to a threshold value of 0.65 mA. Each data point in (c) and (d) is then obtained by calculating the time-averaged switching probability over 100 pulses. All measurements have been performed at 295 K.} 
     \label{Fig_2}
\end{figure*}  

Due to phase inhomogeneities and the unpredictable spatial distribution of different phases, the current in the critical region is inherently probabilistic. If then, a voltage pulse is applied, that initializes the system in the critical state, the measured current is expected to stochastically vary from pulse to pulse. We thus investigate the dynamic response of the system in the critical region, by applying 55 s voltage pulses with an amplitude close to 2 V (1.94 V for L1 and 2.03 V for L2) with a 10 s delay between each pulse. The delay is intended as a reset, to allow the system to cool down and relax back to the initial state. The current response of the system upon the application of two such voltage pulses is shown in Fig. \ref{Fig_1}(d). We observe that both samples display current switching but in a distinctly different manner. In L1, the current switches between two states upon the application of a voltage pulse but remains constant throughout the duration of the pulse itself. In L2, on the other hand, the current switches continuously between multiple states while the voltage pulse is being applied. L1's behavior is reminiscent of a clocked p-bit's operation where an external stimulus initializes the system in a metastable state, after which, it stochastically relaxes back to one of the two stable states \cite{Zink2022, Camsari4}. L2, however, resembles unclocked p-bits where the reduced energy barrier between multiple available states allows the system to stochastically toggle between these states due to thermal fluctuations \cite{Camsari}.

The stochastic switching is more clearly discernible at longer time scales. Figs. \ref{Fig_2}(a) and (b) show the time-evolution of the current upon the application of several voltage pulses. The variation in the switching behavior of both samples, already highlighted in Fig. \ref{Fig_1}(d), is more apparent in these figures. L1 exhibits binary switching between a high-current state (HCS) and a low-current state (LCS). L2, on the other hand, switches between multiple current states but upon closer inspection, it becomes evident that the switching occurs between two current regimes --- a high-current regime (HCR) centered approximately around 0.72 mA and a low-current regime around 0.6 mA. Each current regime appears to consist of multiple current states. Correlating these measurements to the \textit{I-V}s shown in Fig. \ref{Fig_1}(c), we can infer that the origin of the switching is tied to the current spike in the critical region. For both samples, the probability of switching to the higher current state (or regime) can be tuned by modifying the amplitude of the applied pulses. The probability exhibits a sigmoidal dependence on the voltage pulse amplitude, in line with the expected time-averaged output of a p-bit \cite{Camsari, Camsari3}.

\begin{figure}[t]
\centering
     \includegraphics[width=0.483\textwidth]{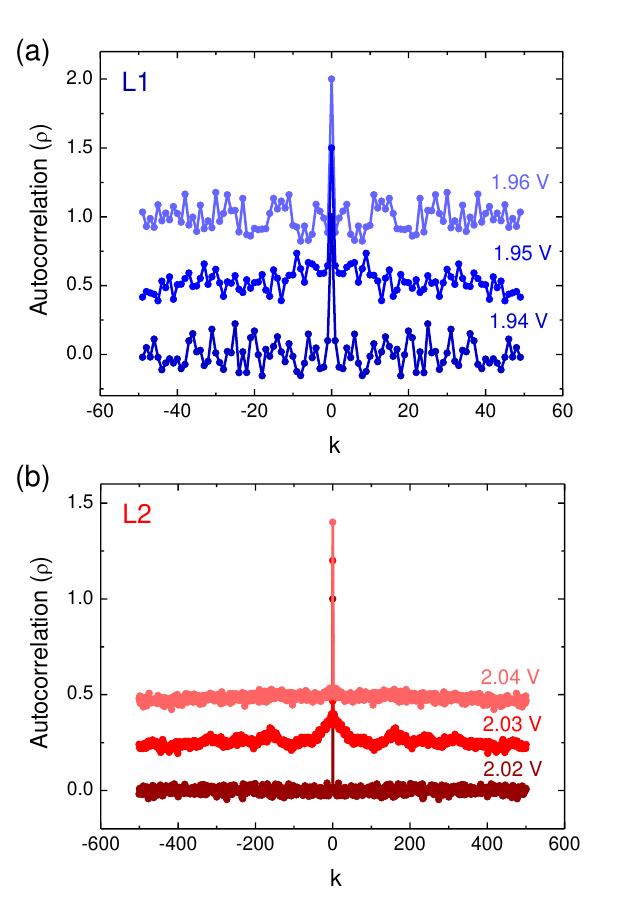}
     \caption{Autocorrelation as a function of the sequence shift, $k$, shown for different voltage pulse amplitudes for (a) L1 and (b) L2. For the sake of clarity, the autocorrelation values of L1 (L2) for 1.95 V (2.03 V) and 1.96 V (2.04 V) have been up-shifted by 0.5 (0.25) and 1.0 (0.5) respectively. The size of the sequences generated by L1 and L2 are different. For L1, each of the 100 applied voltage pulses generates a singular bit, resulting in a 100-bit sequence. For L2, each voltage pulse gives rise to multiple bits, consequently generating a longer bit sequence.} 
     \label{Fig_3}
\end{figure}

To examine whether the generated switching sequence is truly random, we compute the autocorrelation function, $\rho$, defined in Eq. \ref{Eq1}. It compares how similar a given sequence is to itself but with the sequence elements shifted either forward or backward by $k$ positions. $k$ can take on integer values between $-\frac{N-1}{2}$ and $\frac{N-1}{2}$, where $N$ is the total sequence length. 

\begin{equation}
\begin{split}
\rho[k] &= \frac{\sum_{i=0}^{N-k-1} (x_i - \bar{x}) (x_{i+k} - \bar{x})}{\sum_{i=0}^{N-1} (x_i - \bar{x})^2}
\end{split}
\label{Eq1}
\end{equation}

At $k=0$, the sequence is compared to itself and $\rho=\sigma^2$, the variance of the sequence. In this case, $\sigma=1$, since Eq. \ref{Eq1} scales all sequences to unit variance. For a truly random sequence, $\rho$ should be zero for any non-zero value of $k$. Fig. \ref{Fig_3} displays the autocorrelation as a function of the sequence shift for different voltage pulse amplitudes. We observe that the autocorrelation values fluctuate around the zero point for both samples, without any obvious periodicity, indicating a truly stochastic sequence. It is important to keep in mind that for L2, the analyzed sequences represent \textit{inter-regime} current switching. In order to assess whether the \textit{intra-regime} switches are also stochastic, we compute the autocorrelation for the switching sequences within a single current regime, with the results shown in Fig. S4. As expected, the autocorrelation values fluctuate around zero, confirming that the multi-bit sequences produced by L2 are indeed purely stochastic. In addition, the sequences pass applicable tests from the NIST suite \cite{Bassham2010}, further validating their randomness.

Thus, our findings show voltage-tunable stochastic switching in both investigated samples with two key differences: L1 exhibits binary switching while L2 shows multi-bit switching, with L1 operating in a clocked manner and L2 operating in an unclocked manner.

\section{Discussion}

\begin{figure*}[t]
\centering
     \includegraphics[width=0.95\textwidth]{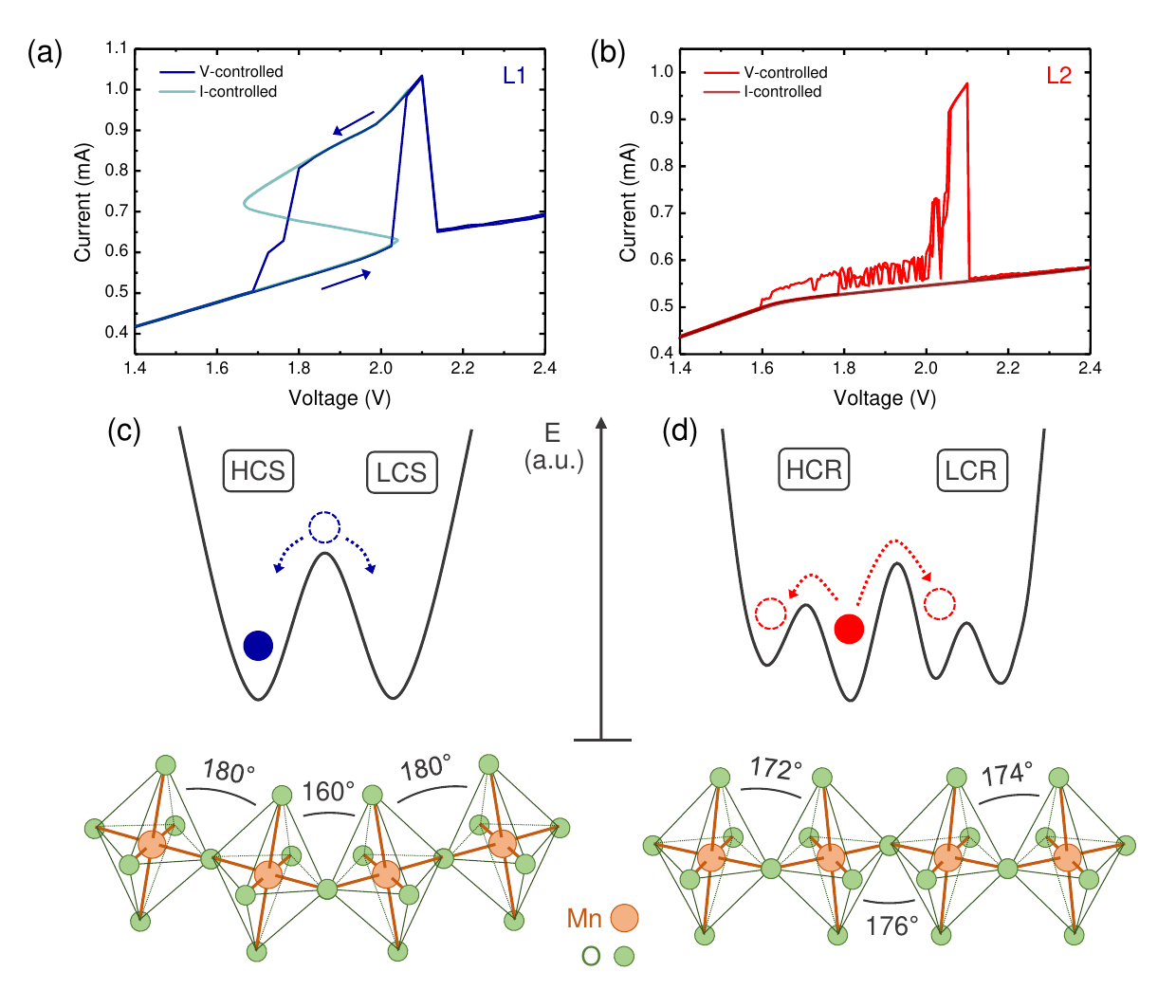}
     \caption{Zoom-in on the critical region in the \textit{I-V} curves of (a) L1 and (b) L2, extracted from Fig. \ref{Fig_1}(c). The proposed energy landscape representing the critical region of (c) L1 and (d) L2. For L1, the system has access to two distinct current states (HCS: high-current state and LCS: low-current state) with a significant barrier between them. For L2, the system can switch between two current regimes (HCR: high-current regime and LCR: low-current regime) with each regime having multiple states. The lower panels of (c) and (d) depict the corresponding Mn-O-Mn bond angle configurations that could explain the proposed landscapes. The bond angle quantities shown are only explanatory examples.} 
     \label{Fig_4}
\end{figure*}

The stochasticity in our samples arises from their behavior in the critical region. Systems undergoing second-order phase transitions transit through a metastable critical region close to the phase transition point. In this region, systems exhibit anomalies such as critical fluctuations of spin, charge and orbital correlations \cite{Imada1998}. Doped manganites are well-known to exhibit coexisting metallic and insulating phases close to the phase transition point \cite{Dagotto2001, Ahn2004, Moreo1999, He2010, Alexandrov2006, Miao2020}. Due to the starkly different resistivity of metallic domains compared to insulating ones, phase coexistence can lead to a highly non-uniform spatial distribution of current and thus, temperature throughout the sample. This scenario is especially true if the MIT is triggered electrically since the sample, while being subject to the voltage sweep, is out of equilibrium. In our system, such inhomogeneities are enhanced by the existence of twin domains, and physically manifest as current nonlinearities in the critical region, as observed in Fig. \ref{Fig_1}(c). It is worthwhile to point out that a critical region with such a clear electrical signature has not been previously reported in other works that have studied the electrically triggered MIT in LSMO \cite{Salev2021, Salev2023a, Salev2023b}. This could be due to the LSMO devices being comparatively small (100$\cross$50 $\mu$m\textsuperscript{2}) and grown on twin-free STO, both factors serving to decrease the vital phase inhomogeneity close to the phase transition. In our case, the four-probe configuration enables the entire 5$\cross$5 mm\textsuperscript{2} resistive network of the film to partake in electrical transport. This collective response, coupled with the intrinsic inhomogeneities arising from the twins, leads to a much richer distribution of coexisting critical phases that can be electrically mapped.

In order to elucidate the reason underlying the two different modes of stochastic switching in L1 and L2, it is essential to inspect the respective critical regions in more detail. Figs. \ref{Fig_4} (a) and (b) zoom in on the \textit{I-V} curves of the two samples, better highlighting the electrical characteristics of the critical region. The binary switching exhibited by L1 is correlated to the abrupt current spike occurring near 2 V in the voltage-controlled \textit{I-V}. The corresponding current-controlled \textit{I-V} shows an S-type negative differential resistance (NDR) which indicates that at a specific voltage, the current can stabilize in different distinct states \cite{Jaman2025}. For L2, the multi-level switching is ascribed to the noisy region preceding the spike, where the current can adopt a multitude of available states.

In Figs. \ref{Fig_4}(c) and (d), we propose energy landscapes describing the critical state that can phenomenologically explain the observed characteristics in both samples. A double-well potential with a significant barrier separating two equivalent energy minima, depicted in Fig. \ref{Fig_4}(c), appropriately describes the critical region of L1. Each minimum here corresponds to one of the current states the system can be in. When a voltage pulse is applied, the system is brought into the critical state and in the process, is momentarily initialized at the top of the potential hill, after which it immediately settles into one of the two states. Due to the energy barrier, the system can only change its state if it is reinitialized to the critical region by another voltage pulse, thereby explaining its clocked behavior. In contrast, the critical state of L2 is best described by a multi-well potential landscape with many competing energy minima separated by low barriers. Each minimum can be assigned to a current state and broadly, two regimes can be defined, as illustrated in Fig. \ref{Fig_4}(d). Once the system is brought to the critical region, it can stochastically toggle between all the available states, aided by thermal fluctuations.

To substantiate the phenomenological interpretation provided by the energy landscapes, the underlying physical mechanism needs to be pinpointed. The primary difference between L1 and L2 lies in the dissimilar spatial configuration of the twins in the two LAO substrates. In LAO, twin domains form to relieve the internal strain arising from the tilting of AlO\textsubscript{6} octahedra in the rhombohedral phase \cite{Hayward2005}. The equilibrium twin domain configuration depends on multiple factors such as aspect ratio of the sample \cite{Scott2022} and temperature \cite{Chrosch1999} and importantly, it varies from one substrate to the other. 

For LSMO, it is well established that octahedral tilting at the interface significantly influences the Mn-O-Mn bond angles, thereby changing the magnetic and transport properties \cite{Liao2016, Moon2014, Li2017}. A deviation of the Mn–O–Mn bond angle from the ideal 180$^{\circ}$ reduces the hopping probability of the \textit{e\textsubscript{g}} electron between adjacent Mn sites, thereby weakening the double-exchange interaction and leading to an increase in the electrical resistance \cite{Dorr2006, Moon2014}. Keeping this in mind, we posit that the twin domains in L1 are sharply angled, leading to more pronounced octahedral tilting and, consequently, greater variations in the Mn-O-Mn bond angles, as depicted in the lower panel of Fig. \ref{Fig_4}(c). Such a configuration results in two distinct current values in the critical state. In addition, it would lead to a lower transition temperature \cite{Garcia-Munoz1996}, which aligns with our observation in Fig. S3. Conversely, the twins in L2 exhibit less acute angles, resulting in a smaller dispersion of bond angles, as illustrated in the lower panel of Fig. \ref{Fig_4}(d). This allows the system to access a broad range of near-equivalent current states in the critical region. 

Our system also has certain limitations. Firstly, the timescale of the clock pulses used for L1 is quite long. The intrinsic timescale of thermodynamic processes like Joule heating depends on multiple factors --- such as the thermal conductivity and the specific heat capacity of a material --- and can be on the order of seconds \cite{Blundell2010}. However, the precise parameter range in which stochastic switching is observed depends on a close interplay between the voltage pulse width, the delay time between pulses and the pulse amplitude. By appropriately tailoring these parameters in our system, stochastic switching can be realized at different timescales. 

Secondly, millimeter-sized samples pose potential challenges when it comes to large-scale circuit integration. Since in our case, the spatially inhomogeneous distribution of current is crucial for stochasticity, it is essential to investigate the effect of downscaling. A promising avenue to explore would be to scale down the LSMO into micrometer-sized devices and study the size-dependent stochastic response in a two-probe configuration.

\section{Conclusion}
In this work, we demonstrate two distinct modes of voltage-tunable stochastic operation in LSMO --- clocked binary switching and unclocked multi-bit switching. By harnessing the rich phase distribution in LSMO near the MIT, and the structural inhomogeneities in LAO, we design a cascade of competing energy states that the system can probabilistically switch between. Autocorrelation analyses show that the switching sequences generated by our systems are purely random, underlining their applicability as true random number generators for cryptography and probabilistic computing. 

\section{Acknowledgments}
I.B. and A.J. acknowledge financial support from the Groningen Cognitive Systems and Materials Center (CogniGron). The authors are thankful to J. G. Holstein, H. H. de Vries and F. H. van der Velde for technical support. I.B., A.J. and T.B. acknowledge K. Y. Camsari and S. Datta for useful discussions. W.Q. acknowledges financial support from the EU-H2020-RISE project MELON (grant no. 872631). This work was realized using NanoLab NL facilities.

\section{Author contributions}
A.J and T.B. conceived the idea. I.B. and A.J. fabricated the devices and performed electrical measurements, along with W.Q and A.G. W.Q. performed the autocorrelation analyses. A.G. carried out the NIST tests. The data was analyzed by I.B and A.J. The results were discussed and their implications agreed upon by all authors. The manuscript was prepared by I.B and T.B with useful input from all authors.

\bibliography{Bibliography}
\end{document}


\title{Supplemental material: Tunable multi-bit stochasticity in \texorpdfstring{La\textsubscript{0.67}Sr\textsubscript{0.33}MnO\textsubscript{3}}{La0.67Sr0.33MnO3}-based probabilistic bits}

\author{Ishitro Bhaduri}
\thanks{Author contributed equally}
 \email{Contact author: i.bhaduri@rug.nl}
\affiliation{Zernike Institute for Advanced Materials, University of Groningen, 9747 AG, Groningen, The Netherlands}
\affiliation{Groningen Cognitive Systems and Materials Center, University of Groningen, 9747 AG, Groningen, The Netherlands}
\author{Azminul Jaman}
\thanks{Author contributed equally}
 \email{Contact author: azminul.jaman@rug.nl}
\affiliation{Zernike Institute for Advanced Materials, University of Groningen, 9747 AG, Groningen, The Netherlands}
\affiliation{Groningen Cognitive Systems and Materials Center, University of Groningen, 9747 AG, Groningen, The Netherlands}
\author{Walter Quiñonez}
\affiliation{Instituto de Nanociencia y Nanotecnología, CONICET-CNEA-CAC, 1650, Buenos Aires, Argentina}
\author{Ayush Gupta}
\affiliation{Zernike Institute for Advanced Materials, University of Groningen, 9747 AG, Groningen, The Netherlands}
\author{Tamalika Banerjee}
 \email{Contact author: t.banerjee@rug.nl}
\affiliation{Zernike Institute for Advanced Materials, University of Groningen, 9747 AG, Groningen, The Netherlands}
\affiliation{Groningen Cognitive Systems and Materials Center, University of Groningen, 9747 AG, Groningen, The Netherlands}
\maketitle

\section{Optical image of the sample surfaces}
\begin{figure}[h]
\centering
     \includegraphics[width=0.6\textwidth]{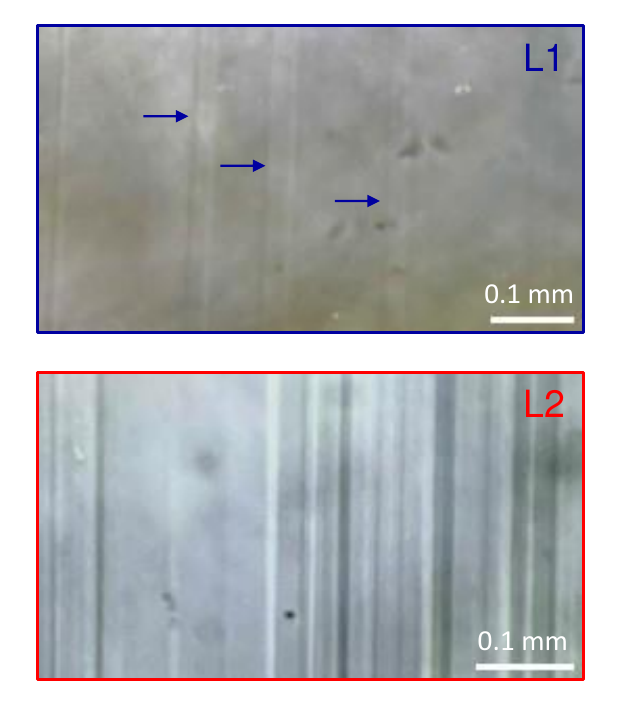}
     \caption{Optical microscopy images of the sample surfaces after LSMO growth. The structure of the twin domains gets transmitted to the surface of the thin film, provided that the film is epitaxial and atomically flat. L1 has a sparse distribution of twins; arrows have been added as visual aids for identifying the domain walls. L2 comparatively, is heavily twinned.}
     \label{Fig_S1}
\end{figure}
\clearpage

\section{Structural characterization of LSMO films}
\begin{figure}[h]
\centering
     \includegraphics[width=\textwidth]{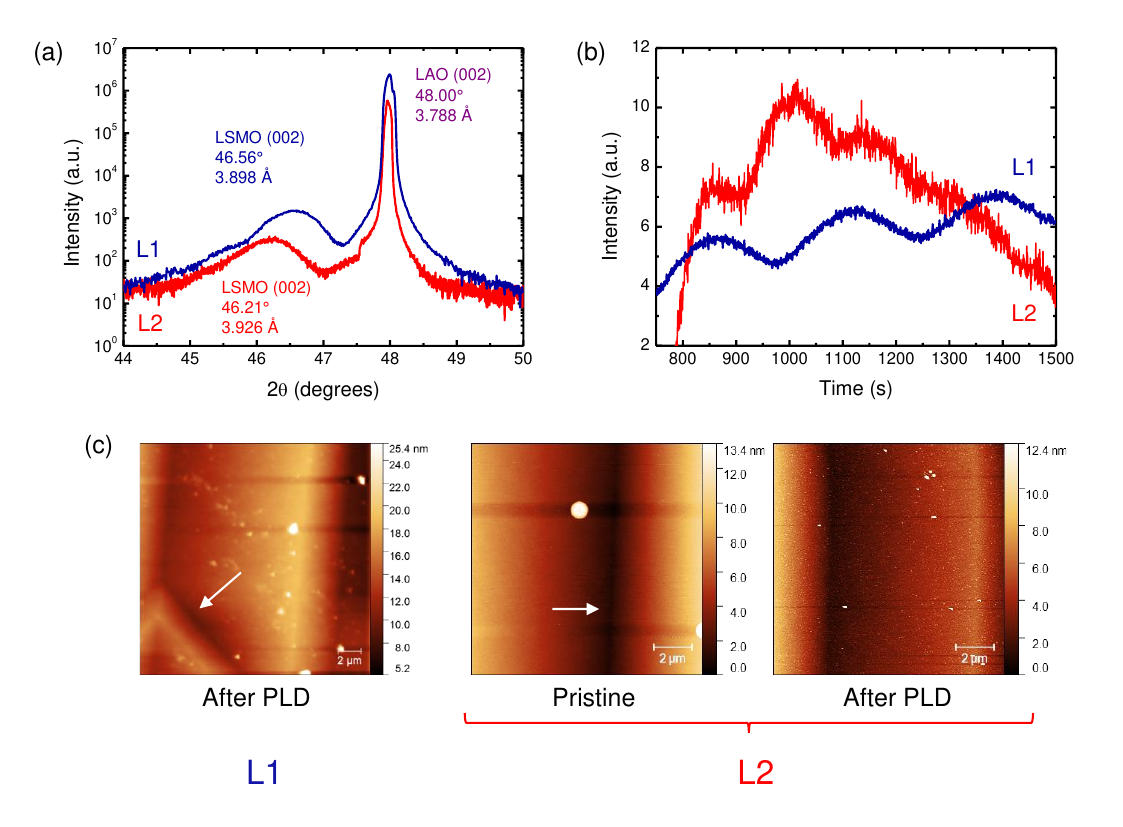}
     \caption{Structural characterization of the LSMO films. (a) XRD 2$\theta$ scan showing the peak belonging to the (002) lattice planes. Both LSMO films are compressively strained in-plane. (b) Time-evolution of the RHEED spot intensity during growth. Intensity oscillations confirm that growth proceeds in a layer-by-layer manner. (c) AFM images of the surface topography before and after PLD growth. Twin domain walls are marked with white arrows for visual clarity.}
     \label{Fig_S2}
\end{figure}

Post-growth AFM scans, presented in Fig. \ref{Fig_S2}(c), reveal that the structure of the twin domains is transmitted to the surface of the LSMO, confirming that the films are epitaxial and atomically flat. To investigate the strain state of the thin films, XRD measurements are performed using a four-axis cradle PANalytical x-ray diffractometer (Cu K$\alpha$, $\lambda$=0.154 nm). 2$\theta$ scans recorded at room temperature, are shown in Fig. \ref{Fig_S2}(a). Intensity peaks corresponding to the (002) lattice planes in LSMO are clearly distinguishable, revealing that both films are compressive-strained in-plane and tensile-strained out-of-plane.

\section{Contact fabrication and electrical measurements}
For electrical measurements, four contacts are fabricated in a van der Pauw geometry. Optical lithography is used to define the circular contacts which are 400 $\mu$m in diameter. Ti/Au constitute the metallic electrodes and are deposited using electron beam evaporation (10 nm Ti and 40 nm Au). The samples are then mounted and wire-bonded to a 44-pin chip carrier which is inserted into a cryostat. All electrical measurements are conducted in the four-probe configuration using a Keithley 2410 source-measure unit.

\section{Determination of \texorpdfstring{\textit{T\textsubscript{MIT}}}{TMIT}}
\begin{figure}[h]
\centering
     \includegraphics[width=0.5\textwidth]{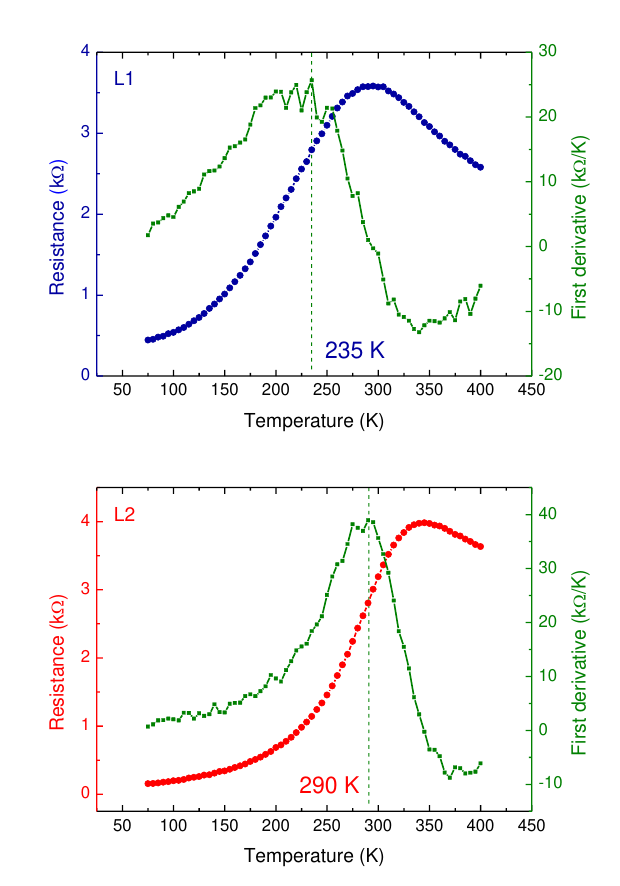}
     \caption{Resistance-temperature curves for both samples plotted alongside the first derivative of the resistance with respect to temperature. Based on the position of the first derivative peak, we determine \textit{T\textsubscript{MIT}} to be 235 K for L1 and 290 K for L2.}
     \label{Fig_S3}
\end{figure}

\section{\textit{Intra-regime} autocorrelation in L2}

\begin{figure}[h]
\centering
     \includegraphics[width=0.8\textwidth]{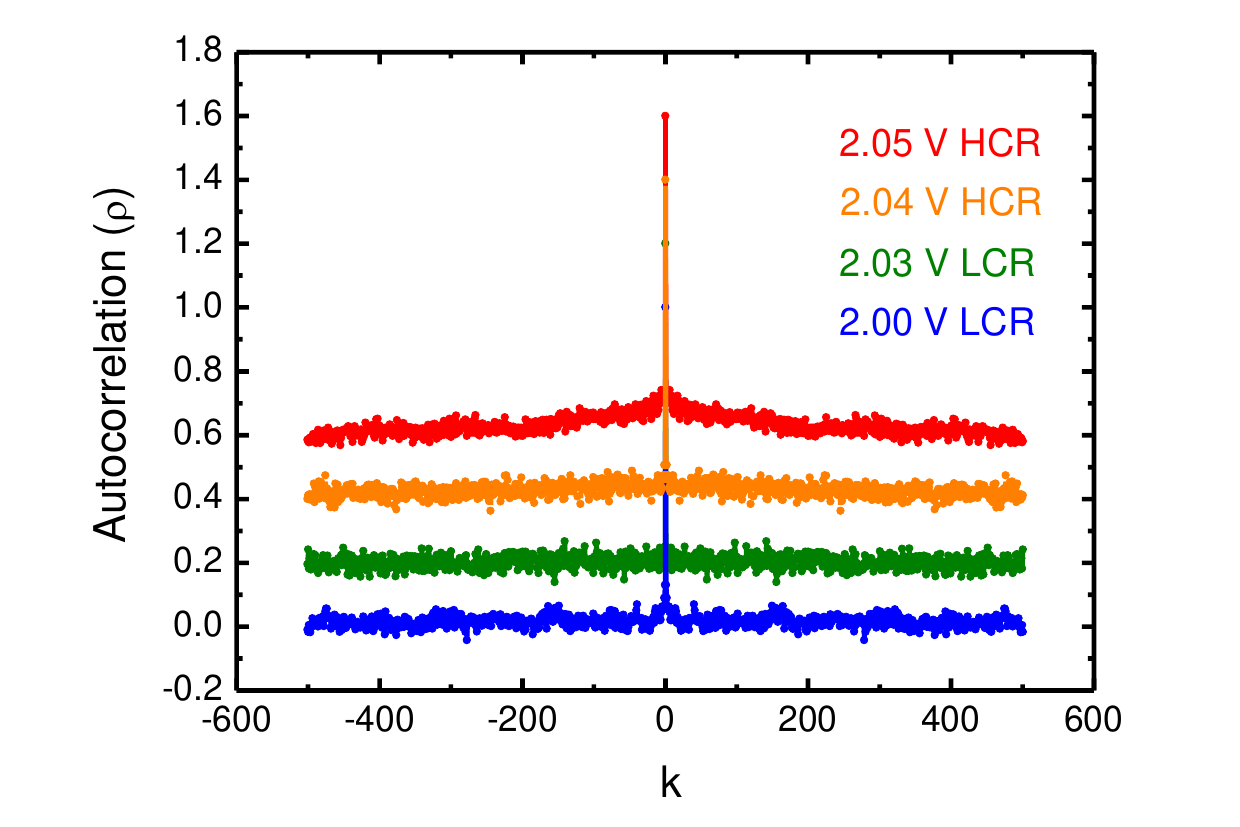}
     \caption{Autocorrelation values as a function of sequence shift for sequences generated by \textit{intra-regime} switching. Some select current regimes are shown here. To ensure clarity, the values for 2.03 V, 2.04 V and 2.05 V have been up-shifted by 0.2, 0.4 and 0.6 respectively. The autocorrelation values randomly fluctuate around zero, confirming that the switches within a single current regime are also stochastic.}
     \label{Fig_S4}
\end{figure}
\clearpage

\section{Randomness assessment using NIST tests}

\begin{table}[h]
\centering
\begin{tabular}{ |p{6cm}|p{3cm}|p{3cm}| }
 \hline
Test & P-value & Result \\
 \hline
 Frequency (Monobit)  & 0.896   &  \textcolor{mygreen}{Random}   \\
\hline
Frequency Test within a Block &  0.849  & \textcolor{mygreen}{Random} \\
\hline
Runs Test & 0.0490 & \textcolor{mygreen}{Random} \\
\hline
Longest Run of Ones in a Block & 0.484 & \textcolor{mygreen}{Random} \\
\hline
\end{tabular}
\caption{Results of the first four randomness tests from the NIST suite. The sequences generated by L2 pass these tests. The P-value, defined by NIST \cite{Bassham2010}, is the probability that a truly random sequence would have produced the test result. In order to pass each test, the P-value needs to be above 0.01.}
\end{table}

To further evaluate the randomness of the sequences produced by L2, we subject them to the statistical tests in the NIST Suite for Random Number Generators for Cryptographic Applications \cite{Bassham2010}. One of the prerequisites of the NIST suite is that the tested sequence should contain an equal number of 1s and 0s with 0.5\% accuracy. This corresponds to the case where the switching probability is 0.5 (at 2.03 V for L2). It is, however, not straightforward to generate a sequence with such an even distribution of 1s and 0s using a physical device. At 2.03 V, the switching probability is, in reality, 0.46. To resolve this, we collect eight independent 520-bit sequences ($S_i$) and perform the nested XOR operation ($\otimes$) as defined by \cite{Fukushima}: $[(S_1 \otimes S_2) \otimes (S_3 \otimes S_4)] \otimes [(S_5 \otimes S_6) \otimes (S_7 \otimes S_8)]$, and successfully tune the probability distribution of 1s and 0s to be within the required (50 $\pm$ 0.5)\% range.

Each NIST test has a recommendation for the minimum size of the input sequences. For the first four tests, this is the lowest, with only a minimum of 100 bits required. In our case, the voltage pulses are quite long, leading to an impractically long acquisition time for generating very large sequences. This is especially true for L1, where each voltage pulse produces a singular bit. Due to these limitations, we have only performed the first four tests from the NIST suite on the sequences generated by L2.
\clearpage

\section{Effect of ambient temperature on the critical region}
\begin{figure}[h]
\centering
     \includegraphics[width=\textwidth]{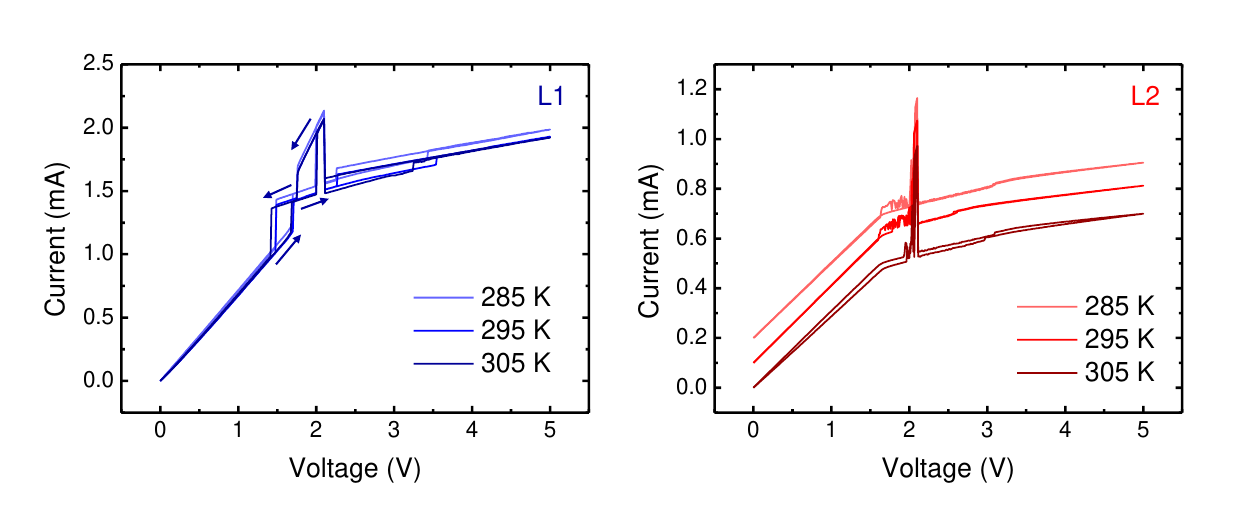}
     \caption{Temperature-dependent current-voltage curves for both samples. For L2, the 295 K and 285 K curves have been upshifted by 0.1 mA and 0.2 mA respectively, for the sake of clarity.}
     \label{Fig_S7}
\end{figure}

We observe that in a 20K range around room temperature, there are no significant changes in the critical region of both samples, indicating that the stochastic functionality in such systems is robust against ambient temperature variations.  

We point out that the critical region of L1 is slightly different in Fig. \ref{Fig_S7} from what has been presented in the main text. This is because a small part of the sample, from the corner, broke during the process of unloading it from the cryostat following the measurements shown in the main text. For the measurements in Fig. \ref{Fig_S7}, a new contact was made close to the broken edge. Thus, the significantly altered geometry affects the current distribution near the MIT, which in turn, modifies the critical region in the \textit{I-V}s.
\clearpage

\bibliography{Bibliography}